\batchmode
\documentstyle[12pt,epsf]{article}
\newcommand\beq{\begin{equation}}
\newcommand\eeq{\end{equation}}
\newcommand{\mt}{m_{\tau}}

\newcommand\tlp{\theta^L_+}
\newcommand\tlm{\theta^L_-}
\newcommand\cp{\cos\tlp}
\newcommand\cm{\cos\tlm}
\newcommand\spp{\sin\tlp}
\newcommand\sm{\sin\tlm}
\newcommand\np{\vec n_+}
\newcommand\nm{\vec n_-}
\newcommand\Pp{\vec p_+}
\newcommand\Pm{\vec p_-}
\newcommand\cph{\cos\varphi}
\newcommand\sph{\sin\varphi}
\newcommand\vd{\vec d}
\newcommand\dmin{\vd_{min}}
\begin{document}
\begin{titlepage}
\title{Semileptonic tau decays,
structure functions, kinematics and polarisation\footnote[2]{
The complete paper, including figures, is
also available via anonymous ftp at
ttpux2.physik.uni-karlsruhe.de (129.13.102.139) as /ttp94-27/ttp94-27.ps,
or via www at http://ttpux2.physik.uni-karlsruhe.de/preprints.html/}
}
\author{
J.~H.~K\"UHN\thanks{Presented at Third Workshop on Tau Lepton
Physics, Montreux, Switzerland, 19 - 22 Sept.\ 1994. Work supported by
BMFT contract 056 KA 93 P.}\\
{\em Institut f\"ur Theoretische Teilchenphysik,} \\
{\em Universit\"at Karlsruhe, }\\
{\em Kaiserstr.12, 76128 Karlsruhe, Germany. }\\[0.3cm]
and\\[0.3cm]
{E.~MIRKES}\\
{\em Physics Department, University of Wisconsin,}\\
{\em Madison, WI 53706, USA }
}
\date{}
\maketitle
\thispagestyle{empty}
\vspace{-12cm}
\begin{flushright}
{\bf TTP 94-27}\\
{\bf hep-ph/9411418}\\
{\bf November 1994}
\end{flushright}
\vspace{9.5cm}
\begin{abstract}
\noindent
{\small
The most general angular distribution
of  two or three meson final states
from semileptonic decays
 $\tau\rightarrow \pi\pi\nu$, $K\pi\nu$,
$\pi\pi\pi\nu$, $K\pi\pi\nu$,
$K\pi\nu$, $KKK\nu$, $\eta\pi\pi\nu,\,\ldots{}$
of polarized $\tau$ leptons
can be characterized by 16 structure functions.
Predictions for hadronic matrix elements, based on CVC and
chiral Lagrangians and their relations to the structure functions
are discussed. Most of them
can be determined in currently ongoing high statistics experiments.
Emphasis of the kinematical analysis is firstly
put on $\tau$ decays in $e^{+}e^{-}$ experiments
where the neutrino escapes detection and the $\tau$ rest frame
cannot be reconstructed.
Subsequently it is shown, how the determination of hadron tracks in
double semileptonic events allows to fully reconstruct the $\tau$
kinematics. The implications for the spin analysis are indicated.
}
\end{abstract}
\end{titlepage}
\section{Structure functions}
\noindent
$\tau$-decays provide an ideal  for
studying strong interaction physics and resonance properties.
Detailed information on the hadronic
charged current for the decay into three pseudoscalar mesons
can be derived in particular from the study of angular distributions.
Consider the semileptonic $\tau$-decay
\beq
\tau(l,s)\rightarrow\nu(l^{\prime},s^{\prime})
+ h_1 + h_2 + h_3 \>,
\label{process}
\eeq
into the pseudoscalar mesons $h_i(q_i,m_i)$ and which is governed by
the  matrix element
\beq
{\cal{M}}=\frac{G}{\sqrt{2}}\,
\bigl(^{\cos\theta_{c}}_{\sin\theta_{c}}\bigr)
\,M_{\mu}J^{\mu}\>,
\label{mdef}
 \eeq
with $G$ the Fermi-coupling constant. The  cosine and the sine of the
Cabbibo angle ($\theta_C$)
in (\ref{mdef}) enter for Cabibbo allowed $\Delta S=0$ and
Cabibbo suppressed $|\Delta S|=1$ decays, respectively.
The leptonic ($M_\mu$) and hadronic ($J^\mu$) currents are given by
\beq
M_{\mu}=
\bar{u}(l^{\prime},s^{\prime})\gamma_{\mu}(g_{V}-g_{A}\gamma_{5})u(l,s)
\eeq
and
\beq
J^{\mu}(q_{1},q_{2},q_{3})=\langle h_1 h_2 h_3
|V^{\mu}(0)-A^{\mu}(0)|0\rangle.
\eeq

$V^\mu$ and $A^\mu$ are the vector and axialvector quark currents,
respectively.
The most general ansatz for the matrix element of the
quark current $J^{\mu}$
is characterized by four formfactors \cite{km1}
\beq
J^\mu
=   V_{1}^{\mu}\,F_{1}
    + V_{2}^{\mu}\,F_{2}
    +\,i\, V_{3}^{\mu}\,F_{3}
    + V_{4}^{\mu}\,F_{4}\label{f1234}\>,
\eeq
with
  \begin{equation}
    \begin{array}{ll}
V_{1}^{\mu}&=q_{1}^{\mu}-q_{3}^{\mu}-Q^{\mu}
\frac{Q(q_{1}-q_{3})}{Q^{2}}\>, \\[2mm]
V_{2}^{\mu}&=q_{2}^{\mu}-q_{3}^{\mu}-Q^{\mu}
\frac{Q(q_{2}-q_{3})}{Q^{2}}\>,\\[2mm]
V_{3}^{\mu}&=\epsilon^{\mu\alpha\beta\gamma}q_{1\,\alpha}q_{2\,\beta}\>
                                            q_{3\,\gamma}
\\[2mm]
V_{4}^{\mu}&=q_{1}^{\mu}+q_{2}^{\mu}+q_{3}^{\mu}\,=Q^{\mu}\>.
    \end{array}
  \end{equation}
The formfactors $F_1$ and $F_2$ ($F_3$) originate from the axial vector
hadronic current (vector current) and lead to spin 1 states,
whereas $F_4$  is due to the spin zero part
of the axial current matrix element.
The formfactors $F_1$  and $F_2$ can be predicted by chiral Lagrangians,
supplemented by informations about resonance parameters.
Parametrizations of the amplitude for the $3\pi$ final states
can be found in \cite{km1,ks,km2}.
In this case the axial formfactors $F_3$ is absent
due to the $G$ parity of the pions.
(The isospin violating transition $\tau\to\nu\pi\omega(\to\pi\pi)$
would, however, give rise to such terms. See \cite{schm})
The formfactor $F_4$ is assumed to be absent as a consequence of PCAC,
an assumption to be tested experimentally.
The $2\pi$ and $3\pi$ decay modes
offer a unique tool for the study of
$\rho,\rho'$ resonance parameters in diffent hadronic enviroments,
competing well with low
enery $e^+e^-$ colliders with energies in the region below 1.7 GeV.
As we will see later, the two body ($\rho$ and $\rho'$) resonance
parameters  can be determined in the $3\pi$ mode by taking
ratios of hadronic structure functions, whereas
the measurement of four structure functions can be used to
put constraints on the $a_1$ parameters.
The decay modes involving
different mesons (for example $K\pi\pi,\,KK\pi$ or $\eta\pi\pi$)
allow for axial and vector current contributions at the same time.
Explicit parametrizations for the form factors in
these decay modes are presented in \cite{dm,dmsw,dproc}.
The vector formfactor $F_3$ is related to the
Wess-Zumino anomaly \cite{wz},
whereas the axial vector form factors are again predicted
by chiral lagrangians. The latter decay modes allow also
for the study of $J^{PC}=0^{-+}$ and
$J^{PC}=1^{++}$ resonances which are not directly accessible
in electron positron annihilation.

Let us now introduce the formalism of the hadronic structure functions.
The differential decay rate  is obtained from
\beq
d\Gamma(\tau\rightarrow \nu_\tau\,3h)=
\frac{G^{2}}{4\mt}\,
\bigl(^{\cos^2\theta_{c}}_{\sin^2\theta_{c}}\bigr)\,
L_{\mu\nu}H^{\mu\nu}
\,d\mbox{PS}\nonumber
 \label{decay}
  \end{equation}
where $L_{\mu\nu}=M_\mu (M_\nu)^\dagger$ and
$H^{\mu\nu}\equiv J^{\mu}(J^{\nu})^{\dagger}$.
The decays are most easily analyzed in the
hadronic rest frame where
$\vec{q}_{1}+\vec{q}_{2}+\vec{q}_{3}=\vec{Q}=0$.
The orientation of the hadronic
system is characterized by
three Euler angles ($\alpha,\beta$ and $\gamma$) as introduced in
\cite{km1,km2}.
In the hadronic rest frame  the product of the hadronic
and the leptonic tensors reduces to the following sum \cite{km1}
\begin{equation} L^{\mu\nu}H_{\mu\nu}=\sum_{X} L_XW_X.
\label{eq1}
\end{equation}
In this system the hadronic tensor $H^{\mu\nu}$ is decomposed into 16
(real) hadronic structure functions $W_{X}$
corresponding to 16 density matrix elements
for a hadronic system in a spin one
[$V_1^{\mu}, V_2^{\mu}, V_3^{\mu}$]
and spin zero state $[V_4^{\mu}]$
(nine of them originate from a pure spin one, one from a
pure spin zero  and six from interference terms).
 The 16 structure functions describe the
dynamics of the three meson decay and depend only on $Q^2$ and
the Dalitz plot variables $s_{i}$.
The factors $L_X$ depend on the Euler angles (which
determine the orientation of the hadronic system),
on the $\tau$ polarization, on the chirality parameter
$\gamma_{VA}$ of the $\tau\nu W$-vertex
and on the total energy of the hadrons in the laboratory frame.
This latter dependence disappears if one considers the $\tau$ decay in
its restframe where it is replaced by the emission angle of
the hadron relative to the $\tau$ spin.
Analytical expressions for the 16 coefficients
$L_X$ were first presented in \cite{km1}.
The formalism can be applied to the case,  where the $\tau$
 rest frame cannot be reconstructed
because of the unknown neutrino momentum and to the case discussed
below, where the full kinematic information is available.
The dependence  of the  coefficients $L_X$ on the  $\tau$ polarization
allows for an improved measurement of the $\tau$ polarization at LEP
[see for example \cite{aleph,paolo,wermes}
and references therein].
\vskip-.5cm
\begin{figure}[tbh]
\begin{center}
\epsfxsize=12cm
\leavevmode
\epsffile[50 0 510 400]{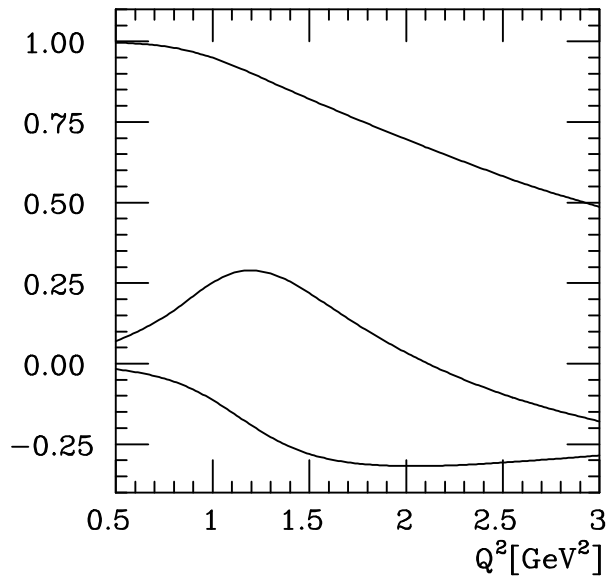}
\end{center}
\vskip -3cm
\caption{Ratio of the spin one hadronic structure functions
$w_C/w_A,w_D/w_A,w_E/w_A$ (from top to bottom) for
$\tau\rightarrow \nu\pi\pi\pi $ as a function of $Q^2$.}
\end{figure}

The hadronic structure functions $W_{X}$
on the other hand contain the full dynamics of the hadronic decay.
The measurement of these structure functions provides
a unique tool for low enery hadronic physics.
They can  be calculated from the components of the hadronic
current $J^{\mu}$ and expressed in terms
of the form factors $F_i$. For brevity
only the results for the pure spin one state are listed.
  \begin{eqnarray}  \hspace{3mm}
W_{A}  &=&   \hspace{3mm}(x_{1}^{2}+x_{3}^{2})\,|F_{1}|^{2}
          +(x_{2}^{2}+x_{3}^{2})\,|F_{2}|^{2}  \nonumber      \\
       && ~~   +2(x_{1}x_{2}-x_{3}^{2})
           \mbox{Re}\left(F_{1}F^{\ast}_{2}\right)
                             \>,      \nonumber \\[3mm]
W_{B}  &=& \hspace{3mm} x_{4}^{2}|F_{3}|^{2}
                                   \nonumber \\[3mm]
W_{C}  &=&  \hspace{3mm} (x_{1}^{2}-x_{3}^{2})\,|F_{1}|^{2}
           +(x_{2}^{2}-x_{3}^{2})\,|F_{2}|^{2} \nonumber       \\
       &&  ~~  +2(x_{1}x_{2}+x_{3}^{2})
           \mbox{Re}\left(F_{1}F^{\ast}_{2}\right)
                                   \nonumber \\[3mm]
W_{D}  &=&  \hspace{3mm}2 [ x_{1}x_{3}\,|F_{1}|^{2}
           -x_{2}x_{3}\,|F_{2}|^{2}         \nonumber      \\
        && ~~  +x_{3}(x_{2}-x_{1})
           \mbox{Re}(F_{1}F^{\ast}_{2})]
                                  \nonumber \\[3mm]
W_{E}  &=& -2x_{3}(x_{1}+x_{2})\,\mbox{Im}\left(F_{1}
                    F^{\ast}_{2} \right) \label{walldef}\\[3mm]
W_{F}  &=&  \hspace{3mm}
          2x_{4}\left[x_{1}\,\mbox{Im}\left(F_{1}F^{\ast}_{3}\right)
                     + x_{2}\,
                     \mbox{Im}\left(F_{2}F^{\ast}_{3}\right)\right]
                                  \nonumber \\[3mm]
W_{G}  &=&- 2x_{4}\left[x_{1}\,\mbox{Re}\left(F_{1}F^{\ast}_{3}\right)
                     + x_{2}\,
                     \mbox{Re}\left(F_{2}F^{\ast}_{3}\right)\right]]
                                   \nonumber \\[3mm]
W_{H}  &=& \hspace{3mm}
      2x_{3}x_{4}\left[\,\mbox{Im}\left(F_{1}F^{\ast}_{3}\right)
                     -\,\mbox{Im}\left(F_{2}F^{\ast}_{3}\right)\right]
                                   \nonumber \\[3mm]
W_{I}  &=&- 2x_{3}x_{4}\left[\,\mbox{Re}\left(F_{1}F^{\ast}_{3}\right)
                     -\,\mbox{Re}\left(F_{2}F^{\ast}_{3}\right)\right]
                                 .  \nonumber
\end{eqnarray}
The remaining structure functions originating
from a possible (small) contribution
of a spin zero state are presented  in [1].
The variables $x_i$ are defined by
$
x_{1}= V_{1}^{x}=q_{1}^{x}-q_{3}^{x},\,
x_{2}= V_{2}^{x}=q_{2}^{x}-q_{3}^{x},\,
x_{3}= V_{1}^{y}=q_{1}^{y}=-q_{2}^{y},\,
x_{4}= V_{3}^{z}=\sqrt{Q^{2}}x_{3}q_{3}^{x},\,
$
where $q_i^{x}$ ($q_i^{y}$) denotes
the $x$ ($y$) component of the momentum of
meson $i$ in the hadronic rest frame.
They can easily be expressed in terms of $s_1$, $s_2$ and $s_3$
\cite{km1,km2}.
The structure functions can be extracted by taking suitable moments
with respect to an appropriate product of sine or cosine of
two Euler angles. An alternative method to extract the structure
functions has been suggested in \cite{wermes}
where a direct fit to the expressions (\ref{walldef}) was performed.

As an example, we will now present numerical results for the
non vanishing structure functions $W_A, W_C, W_D$ and $W_E$
in the $3\pi$ decay mode.
Figure 1 shows predictions for the structure function ratios
$w_C/w_A, w_D/w_A$ and $w_E/w_A$ as a function of $Q^2$, where
we have integrated over the Dalitz plot variables $s_i$ [The
integrated structure functions are denoted by lower case letter $w_X$].
The results are based on the same parametrization of
the formfactors as those used in \cite{km1}. Although information
on the resonance parameters in the  two body decays is lost
by integrating over $s_1$ and $s_2$, interesting structures
\footnote{More constraints on the
two body resonances can be obtained by analyzing the full
dependence on $Q^2$ and $s_i$, which should be accessible with the
present  high statistic experiments.}
are observed nevertheless.
One observes that all normalized structure functions are  sizable.
$w_{C}/w_{A}$ approaches its maximal value $1$ for small $Q^{2}$.
The ``time reversal'' invariant ratios $w_C/w_A$ and $w_D/w_A$ reach up
to unity, the ``non-invariant'' ratio $w_E/w_A$ which is sensitive
towards phases is significantly smaller.
Note, that the dependence on the $a_1$ mass and width parameters
cancel in the ratio $w_X/w_A$ in fig.~1.
\vskip-.5cm
\begin{figure}[tbh]
\begin{center}
\epsfxsize=13cm
\leavevmode
\epsffile[0 0 510 400]{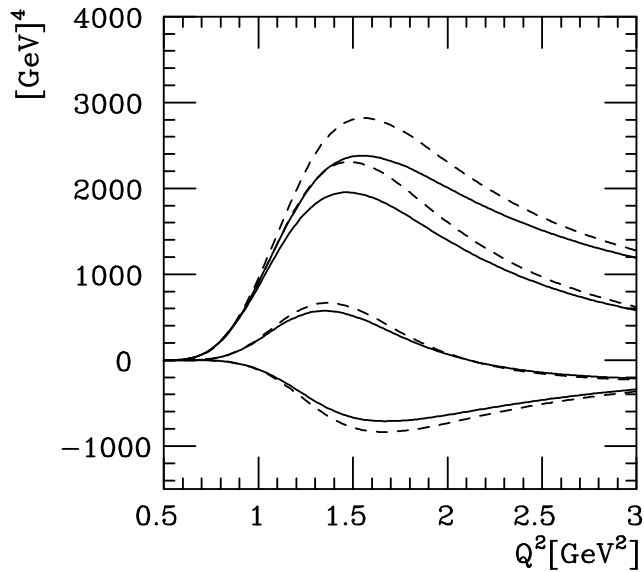}
\end{center}
\vskip-3cm
\caption{Spin one hadronic structure functions
$w_A,w_C,w_D,w_E$ (from top to bottom) for
$\tau\rightarrow \nu\pi\pi\pi$
as a function of $Q^2$.
Results are shown for two sets of $a_1$ parameters:
$m_{a_1}=1.251$ GeV, $\Gamma_{a_1}= 0.599$ GeV (solid)
and
$m_{a_1}=1.251$ GeV, $\Gamma_{a_1}= 0.550$ GeV (dashed) }
\end{figure}
On the other hand, the $Q^2$ distributions of the structure functions
$w_{A,C,D,E}$ presented in fig.~2 are
very sensitive to the $a_1$ parameters.
As an example, fig. 2 shows predictions for the structure functions,
where two different values for the $a_1$ width have been used.
Therefore, the ratios in fig.~1 can be used to fix the model dependence
in the two body resonances, whereas the structure functions themselves
impose rigid contraints on the $a_1$ parameters.
Through measurements of the $W_X$ it is therefore possible
to determine the amplitudes in much more detail than through
rate measurements alone.

The technique of structure functions also allow for a model independent
search for a spin zero component in the hadronic current.
Such a contribution would lead to  the six forementioned
additional structure functions
originating from the interference with the (large)
spin one contributions\cite{km1}.

The analysis of angular distributions of the hadronic final states
allows to determine not only the properties of the hadronic current, it
is also an important tool for the determination of the $\tau$
polarization and the structure of the $\tau\nu W$ coupling. This holds
true for single pion final states, but is equally valid for decays into
two\cite{KW,hagiwara,rouge} or three\cite{KW,rouge,km1,km2} pions, if
the angular distributions are fully explored. As demonstrated in
\cite{davier,aleph,paolo} the $L_CW_C$ term is particularly sensitive
towards the $\tau$ polarization, whereas the $L_EW_E$ term determines
the helicity of the $\tau$-neutrino.

A detailed discussion of the matrix elements for
the decay modes involving different pseudoscalar mesons
[$K\pi\pi\nu,\, KKK\nu,\, \eta\pi\pi\nu$]
 together with predictions
for the corresponding structure functions and angular distributions
is presented in \cite{dm}.
In this case, all 9 structure functions in (\ref{walldef}) are
nonvanishing because of the interference of the anomaly with
the axial vector contributions.
The analysis of these distributions would  allow to test
the underlying  hadronic physics, and to separate for example
the contributions from
the axial and the vector (Wess-Zumino anomaly) current.
It is thus possible to confirm  (not only qualitatively)
the presence of the Wess-Zumino anomaly in the decay modes
 $\tau\rightarrow \nu K\pi\pi$ and
$\tau\rightarrow \nu K\pi\pi$.

In \cite{dm1} the technique of the structure functions
has been  extended to the $\tau\rightarrow\nu\omega\pi$ decay mode.
This allows to test the model for the hadronic matrix element,
which involves both a vector and a second class axial vector current.
\begin{figure}[htb]
\begin{center}
\leavevmode
\epsfxsize=14cm
\epsffile[0 0 425 179]{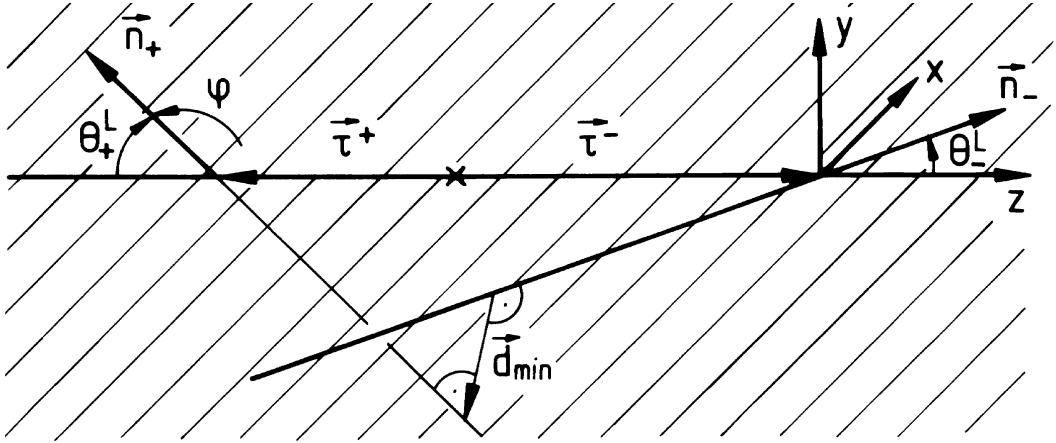}
\end{center}
\vskip -.5cm
\caption{
Kinematic configuration indicating the relative orientation of the
hadronic tracks, the $\tau$ directions and the vector $\dmin$.
\label{Kin}}
 \end{figure}

\section{Tau kinematics}

\noindent
In all experimental and theoretical analysis of $\tau$ decays it
has been implicitly assumed
that the $\tau$ direction and its restframe cannot be fully
reconstructed and appropriately averaged distributions are considered.
However, as shown in \cite{impact}, impact parameter measurements allow
to fully reconstruct the $\tau$ direction.

In events where both leptons decay semileptonically and where all
hadron momenta are determined, the original $\tau$ direction can be
reconstructed up to a twofold ambiguity \cite{Tsai,KW,davier},
which can be resolved with the help of vertex detectors
employed in present experiments.  Several possibilities may
arise:

{\em i.)}  If the beam spot is large compared to the typical impact
parameter, the production vertex is unknown.  Let us assume that
both $\tau$ decay into one charged hadron each and that both
charged tracks can be measured with high precision.  The
direction $\vec d_{min}$
of the minimal distance between the two nonintersecting
charged tracks (Fig.1) resolves the ambiguity and introduces two
additional constraints that can be used to reduce the measurement
errors.  The $\tau^+$ and $\tau^-$ decay points and their original
directions of flight are then uniquely determined.

{\em ii.)}  Precise knowledge of the beam axis (corresponding to a
beam spot of negligible size) leads to a further constraint
resulting from the requirement that the reconstructed $\tau$ axis
and the beam axis intersect.  If the production point would be
known in addition then the momenta plus one charged track from the decay
of only one $\tau$ would allow to reconstruct the event
and double semileptonic decays
would lead to a large number of additional constraints.

{\em iii.)}  If one (or both) $\tau$ decay into several hadrons and
if all momenta and two tracks (one from each side) are measured,
the same reconstruction can be performed.

Let us first consider case {\em i)} where all relevant aspects can be
explained most clearly.  The angles $\theta^L_{\pm}$ between the
$\tau^\pm$ and the  hadron $h^\pm$
directions respectively as defined in the lab frame are
given by the energies of $h^+$ and $h^-$ \cite{KW}:
\beq
\cm = \frac{\gamma x_- - (1+r^2_-)/2\gamma}
                      {\beta\sqrt{\gamma^2x_-^2-r_-^2}}
\eeq
\[
\sm =
    \sqrt{\frac{(1-r_-^2)^2/4 - (x_--(1+r^2_-)/2)^2/\beta^2}
      {\gamma^2x_-^2-r_-^2}}
\]
\beq
x_-=E_{h^-}/E_\tau \hspace{2cm}   r_-=m_{h^-}/m_\tau
\eeq
and similarly for $\cp$ and $\spp$. The velocity
$\beta$, and the boost factor $\gamma$ refer to the $\tau$ in the lab
frame.

The original $\tau^-$ direction must therefore lie on the cone of
opening angle $\tlm$ around the direction of $h^-$ and on the cone of
opening angle $\tlp$ around the reflected direction of $h^+$.  The
extremal situation where $\tlp$ or $\tlm$ assume the values 0 or $\pi$,
or where the
two cones touch in one line, lead to a unique solution for the $\tau$
direction.  In general a twofold ambiguity arises as
is obvious from this geometric argument. The
cosine of the relative azimuthal angle $\varphi$ between the directions
of $h^+$ and $h^-$ denoted by $\np$
and $\nm$ can be calculated from the momenta
and energies of $h^+$ and $h^-$ as follows:  In the coordinate frame
(see Fig.1) with the $z$ axis pointing along the direction of $\tau^-$
and with $\nm$ in the $xz$ plane and positive $x$ component
\beq
\frac{\Pm}{|\Pm|}\equiv\nm=
\left(\begin{array}{c} \sm    \\0       \\\cm     \end{array}\right)
\nonumber
\eeq
\beq
\frac{\Pp}{|\Pp|}\equiv\np=
\left(\begin{array}{c}\spp\cph \\ \spp\sph\\-\cp
\end{array}\right)
\eeq
and $\cph$ can be determined from
\beq
\nm\np=-\cm \cp + \sm \spp\cph
\eeq
The well-known twofold ambiguity in $\varphi$ is evident from
this formula.

Additional information can be drawn from the precise
determination of tracks close to the production point.
Three-prong decays allow to reconstruct the decay vertex and the
ambiguity can be trivially resolved.

However, also single-prong events may serve this purpose.  Let us
first consider decays into one charged hadron on each side.
Their tracks and in particular the vector $\dmin$ of closest
approach
(Fig.1) can be measured with the help of microvertex detectors.
The vector pointing
from the $\tau_-$  to the $\tau_+$ decay vertex
\beq
\vec d\equiv \vec \tau_+ -\vec\tau_-= -l
\left(\begin{array}{c}0\\0\\1\end{array}\right)
\eeq
is oriented  by definition into the negative $z$ direction
($l>0)$.
The vector $\dmin$ can on the one hand be measured, on the other hand
calculated from $\vec d$, $\np$ and $\nm$:
\begin{eqnarray}\label{eqdmin}
\lefteqn{\dmin=\vd\, + \,     [(\vd\np\,\np\nm - \vd\nm)\nm}\\
&& + (\vd\nm\,\np\nm - \vd\np)\np]
/(1-(\nm\np)^2)
\end{eqnarray}
The sign of the projection of $\dmin$ on $\np\times\nm$ then
determines the sign of $\varphi$ and hence resolves the
ambiguity.
\beq
\dmin(\np\times\nm) = l \spp\sm\sph
\eeq
The length of the projection determines $l$ and hence
provides a measurement of the lifetimes
of
$\tau_+$ plus $\tau_-$.
Exploiting the fact that $\vd\nm=-l\cm$ and $\vd\np=l\cp$ the direction
of $\vd$ can be geometrically constructed by inverting (\ref{eqdmin}):
\begin{eqnarray}
\lefteqn{ \vd/l=\dmin/l \, - \, [(\cp \, \np\nm + \cm)\nm+}\\
&&  (-\cm \, \np\nm - \cp)\np] / (1-(\nm\np)^2)\nonumber
\end{eqnarray}
In addition two constraint equations may be
derived by comparing $\dmin/l$
as calculated from the $h_+$ and $h_-$ tracks with
the direction calculated from (\ref{eqdmin})
with the help of $\tlp$, $\tlm$ and $l$.
These might be used to constrain the events even in cases where initial
state radiation distorts the simple kinematics described above.

As stated before, the locations of both $\tau_+$ and $\tau_-$
decay vertices
in space are then fixed.  If the beam axis is known with high
precision (high compared to the decay length $l$) the lines between
the two decay vertices and the beam axis intersect, providing one
additional constraint.

The generalization of this method to decays into multihadron
states with one or several neutrals is straightforward: $\tlp$,
$\tlm$ and $\cos\varphi$ are
fixed by the hadron momenta as stated above.  Only one of the two
solutions for $\varphi$ is then compatible with $\dmin$
measured directly with the help of vertex detectors.

\section{\em $\tau$-polarization}

\noindent
As discussed in chapter 1, the technique of structure functions serves
as a usefull tool to analyse the hadronic current and the
$\tau$-polarization. Once the $\tau$ restframe is reconstructed one may
also proceed in a more direct way which allows to exploit maximally the
spin information on an event by event basis.

The current $J_\mu$
depends generically on all hadron momenta. In \cite{KW} it
has been demonstrated that the direction of the $\tau$-spin
in the $\tau$-restframe in each event is given by
\beq
\vec h = \vec H/\omega
\eeq
with
\beq
\omega=P^\mu(\Pi_\mu - \Pi^5_\mu); ~~~~~
\vec H=\mt(\vec\Pi^5-\vec\Pi)
\eeq
\begin{eqnarray}
\Pi_\mu&=&2[(J^*\cdot N) J_\mu + (J\cdot N)J^*_\mu-(J^*\cdot J) N_\mu];
\nonumber\\
\Pi^5_\mu&=&
2\,{\rm Im}\, \epsilon_\mu^{\phantom{\mu}\nu\rho\sigma}
J_\nu^*J_\rho N_\sigma.
\end{eqnarray}

The momenta $P$ and $N$ refer to the $\tau$ and its neutrino.
The $V-A$ structure of the lepton current and a massless neutrino have
been  assumed for simplicity.

For the decay into a single pion $J_\mu\propto Q_\mu$ and
the direction of the vector $\vec h$ is
identical to that of the momentum of the pion, its  normalization
equal to one. For the decay into several pions the current $J_\mu$ and
consequently the direction of $h$ will depend on all pion momenta. It
is, however, a simple exercise  to demonstrate that
\beq
\vec h^2 = 1\,
\eeq
also in the general case. This follows directly from
\beq
(\Pi_\mu - \Pi_\mu^5) (\Pi^\mu - \Pi^{5\mu}) = 0
\eeq
and is a direct consequence of the $V-A$ coupling at the lepton vertex
and of the fact that no spin summation has been performed  for the
hadronic system.
Also decay modes with kaons or $\eta$'s only have maximal
analysing power, corresponding to $|\vec h|=1$.

The argument is in particular also applicable for $\tau\to
\nu\pi\omega(\to 3 \pi)$ if all four pion momenta are measured and the
full matrix element including also the $\omega\to3\pi$ decay is used for
the hadronic current. The argument would fail, if the spin of the
$\omega$ meson would be averaged. Correspondingly, $|\vec h|\neq 1$ for
leptonic decay modes.

However, one important assumption has been made tacidly throughout: The
form of the current must be known for the analysis of multi-hadron
states. For 2 pions this form is essentially fixed, for 3 pions it is
highly plausible. For multi-meson states, in particular those including
kaons, a carefull test of the functional form of $J$ is madatory.

{\em To summarize:}
Semileptonic decay modes of the $\tau$-lepton carry important
information on hadron physics and $\tau$-properties. Measurements of the
structure functions are a convenient tool to constrain models and to
determine the hadronic current in a model-independent way.
Given sufficiently precise  vertex detectors, double-semileptonic events
can be fully reconstructed. Once the model for the hadronic current is
reliable, the $\tau$ spin can be measured in
semileptonic decays on an event by event basis.

\end{document}